\documentclass[aps]{revtex4}

\usepackage{amsmath}
\usepackage{graphicx}

\usepackage{epsfig}

\begin{document}
\begin{flushright}
\global\long\def\ov#1{\overline{#1}}
 \global\long\def\un#1{\underline{#1}}
 \global\long\def\wt#1{\widetilde{#1}}
 \global\long\def\wh#1{\widehat{#1}}
 \global\long\def\vb#1{\boldsymbol{#1}}
 \global\long\def\pd#1#2{\frac{\partial#1}{\partial#2}}
 \global\long\def\fd#1#2{\frac{\delta#1}{\delta#2}}
 \global\long\def\btimes{\boldsymbol{\times}}
 \global\long\def\bdot{\boldsymbol{\cdot}}
 \global\long\def\bhat{\widehat{\mathsf{b}}}
 \global\long\def\exd{\mathsf{d}}
 \global\long\def\cont#1{\iota_{#1}}
April 6, 2010
\par\end{flushright}

\title{Perturbation analysis of trapped-particle dynamics in axisymmetric dipole geometry}

\author{F.-X.~Duthoit, A.~J.~Brizard\footnote{Permanent address: Department of Physics, Saint Michael's College, Colchester, VT 05439, USA}, Y.~Peysson,
and J.~Decker}

\affiliation{CEA, IRFM, F-13108, Saint-Paul-lez-Durance, France}

\begin{abstract}
The perturbation analysis of the bounce action-angle coordinates $(J,\zeta)$ for charged particles trapped in an axisymmetric dipole magnetic field is presented. First, the lowest-order bounce action-angle coordinates are derived for deeply-trapped particles in the harmonic-oscillator approximation. Next, the Lie-transform perturbation method is used to derive higher-order anharmonic action-angle corrections. Explicit expressions (with anharmonic corrections) for the canonical parallel coordinates $s(J,\zeta)$ and $p_{\|}(J,\zeta)$ are presented, which satisfy the canonical identity $\{ s,\;
p_{\|}\}(J,\zeta) \equiv 1$. Lastly, analytical expressions for the bounce and drift frequencies (which include anharmonic corrections) yield excellent agreement with exact numerical results. 
\end{abstract}

\maketitle


\section{Introduction}

The dynamics of magnetically-confined charged particles exhibits three
orbital time scales \cite{Tao_07} associated with the fast
gyromotion, the intermediate bounce motion, and the slow drift motion around, along, and across magnetic-field lines, respectively.
To lowest order, the fast gyromotion involves circular motion of a
charged particle (with mass $m$ and charge $q$) in the plane transverse
to a magnetic field line. The radius of gyration (or gyroradius) $\rho_{\mathrm{g}}=v_{\bot}/|\omega_{\mathrm{g}}|$
is defined as the ratio of the magnitude of the particle's perpendicular
velocity $v_{\bot}=|\mathbf{v}_{\bot}|$ and the gyrofrequency $\omega_{\mathrm{g}}=qB/mc$.
When the gyroradius $\rho_{\mathrm{g}}$ is much smaller than the
magnetic nonuniformity length-scale $L_{\mathrm{B}}$ (i.e., $\epsilon\equiv\rho_{\mathrm{g}}/L_{\mathrm{B}}\ll1$),
the gyromotion is nearly circular and the action variable $\mu B/\omega_{\mathrm{g}}$
associated with gyromotion is an adiabatic invariant (here $\mu\equiv mv_{\bot}^{2}/2B$
denotes the magnetic moment).

The slower bounce and drift motions of a guiding-center are described by the guiding-center Hamiltonian 
\begin{equation}
H_{\mathrm{g}} \;=\; \frac{p_{\|}^{2}}{2m} \;+\; \mu B
\label{eq:Ham_gc}
\end{equation}
and the guiding-center phase-space Lagrangian 
\begin{equation}
\Gamma_{\mathrm{g}} \;=\; \left(\frac{q}{\epsilon c}\mathbf{A} \;+\; p_{\|}\bhat\right)\bdot\exd\mathbf{X} \;-\; H_{\mathrm{g}}\;\exd t,
\label{eq:psl_gc}
\end{equation}
where $\mathbf{A}$ denotes the vector potential generating the magnetic field $\mathbf{B}\equiv\nabla\btimes\mathbf{A}$, $p_{\|} = mv_{\|}$ denotes the guiding-center's parallel momentum, and the term $\mu B$ plays the role of a spatial potential in which the guiding-center moves. We note that the ignorable gyromotion action-angle pair $\epsilon\,(mc/q)\,\mu\,\exd\zeta_{\rm g}$ has been omitted in the guiding-center phase-space Lagrangian 
\eqref{eq:psl_gc} since it plays no dynamical role in bounce-center dynamics. 

Guiding-center dynamics takes place in a four-dimensional phase space \cite{RGL_83}, with the magnetic moment $\mu$ treated as an invariant and the gyroangle $\zeta_{\rm g}$ treated as an ignorable angle. The magnetic moment $\mu$ and the total energy $\mathcal{E}= p_{\|}^{2}/2m+\mu B$ are thus constant parameters for a given guiding-center orbit. By using the parallel spatial coordinate $s$ along a magnetic-field line labeled by the potentials 
$(\alpha,\beta) \equiv (y^{1}, y^{2})$, for which $\mathbf{B}\equiv\nabla\alpha\btimes\nabla\beta$ and $\bhat\equiv\partial\mathbf{X}/\partial s$, the guiding-center Lagrangian \eqref{eq:psl_gc} becomes 
\begin{equation}
\Gamma_{\mathrm{g}} \;=\; \frac{q}{\epsilon c}\;\alpha\,\exd\beta \;+\; p_{\|}\left(\exd s\;+\; b_{i}\,\exd y^{i} \right) \;-\; H_{\mathrm{g}}\;
\exd t,
\label{eq:psl_alphabeta}
\end{equation}
where $\mathbf{A}\equiv\alpha\nabla\beta$ and $\bhat \equiv \nabla s + b_{i}\nabla y^{i}$
have been substituted into Eq.~\eqref{eq:psl_gc}, and summation over repeated indices is henceforth implied (unless stated otherwise).
The non-vanishing covariant components $b_{i}\equiv\bhat\bdot\partial\mathbf{X}/\partial y^{i}$ guarantee that, using $\bhat\bdot\nabla y^{i} \equiv 0$, the magnetic curvature $\bhat\bdot\nabla\bhat=(\bhat\bdot\nabla b_{i})\nabla y^{i}$ is non-vanishing. In addition, we note that the covariant components
may also be expressed as $b_{i} = -\nabla s\bdot\nabla y^{i}/|\nabla y^{i}|^{2}$ (no summation), and thus the magnetic coordinates $(\alpha,\beta,s)$
are in general non-orthogonal (if at least one component $b_{i}$ is non-vanishing and we assumed $\nabla\alpha\bdot\nabla\beta \equiv 0$).

The present work draws its motivation from some practical aspects
of the bounce-center phase-space transformation used to obtain the
bounce-center phase-space Lagrangian \cite{RGL_82a,AJB_00,Cary_Brizard}
\begin{equation}
\ov{\Gamma}_{\mathrm{b}} \;\equiv\; \frac{q}{\epsilon c}\;\ov{\alpha}\,\exd\ov{\beta} \;+\; \ov J\;\exd\ov{\zeta} \;-\; \ov H_{\mathrm{b}}\;\exd t,
\label{eq:psl_bc}
\end{equation}
where the coordinates $(\ov{\alpha},\ov{\beta})$ denote the
bounce-center (e.g., banana-center in tokamak geometry) coordinates, while
the bounce-center action-angle coordinates $(\ov J,\ov{\zeta})$ form
a canonical pair (where the bounce angle $\ov{\zeta}$ is ignorable
and the bounce action $\ov J$ is an adiabatic invariant). In Eq.~\eqref{eq:psl_bc}, the bounce-center
Hamiltonian $\ov{H}_{\rm b}(\ov{\alpha}, \ov{\beta}, t; \ov{J}, \mu)$ is a function of the bounce-center coordinates $(\ov{\alpha},\ov{\beta})$ and depends parametrically on the bounce-action $\ov{J}$ and the guiding-center magnetic moment $\mu$. Although the
bounce-center phase-space transformation \cite{RGL_82a} follows the same Lie-transform
perturbation approach \cite{RGL_82b} used to derive the guiding-center phase-space
Lagrangian \eqref{eq:psl_gc}, there are differences that are investigated
in the present work.

The remainder of this paper is organized as follows. In Sec.~\ref{sec:bc_general},
we review the theory of bounce-center dynamics in general magnetic
geometry. By using magnetic coordinates $(\alpha,\beta,s)$, the lowest-order 
guiding-center parallel dynamics is clearly separated from the slower 
guiding-center drift motion. From the parallel dynamics, the
lowest-order bounce action-angle coordinates $(J,\zeta)$ are introduced
and the breaking of the lowest-order bounce action by the slow guiding-center
drift motion is discussed. Next, the general formulation of the bounce-center
phase-space transformation is presented and the crucial role played
by the canonical parallel-dynamics relation $(s,p_{\|})\rightarrow(J,\zeta)$
is discussed. While the explicit proof of this canonical relation is not known 
in general magnetic geometry, it is investigated in axisymmetric magnetic dipole
geometry in Sec.~\ref{sec:gc_dipole}. In this simple magnetic geometry,
the lowest-order bounce action is in fact an exact guiding-center
invariant (which partially negates the need for the bounce-center
transformation). We are thus able to explore, in Sec.~\ref{sec:bc_deep},
the canonical parallel-dynamics relation for trapped particles,
first, in the deeply-trapped (harmonic-oscillator) approximation and, then, by considering
anharmonic corrections. Lastly, in Sec.~\ref{sec:summary}, we summarize
our work and discuss its application to more complex magnetic geometries
(e.g., axisymmetric tokamak magnetic geometry).

\section{\label{sec:bc_general}Bounce-center Dynamics in General Magnetic Geometry}

The bounce-center dynamical equations \cite{RGL_82a} describe the adiabatic motion of magnetically-confined charged particles in a nonuniform magnetic field, in which the fast bounce-motion time scale in the four-dimensional guiding-center motion have been asymptotically decoupled from the two-dimensional slow drift motion across the field lines.

\subsection{Four-dimensional guiding-center dynamics}

We begin our general discussion of bounce-center dynamics with guiding-center Hamiltonian dynamics in four dimensional phase space based on the guiding-center phase-space Lagrangian \eqref{eq:psl_alphabeta}. The Euler-Lagrange equations obtained from the guiding-center phase-space
Lagrangian \eqref{eq:psl_alphabeta} are \cite{Cary_Brizard} 
\begin{eqnarray}
\dot{y}^{i} & = & \eta^{ij}\frac{c\epsilon}{q\mathcal{J}} \left[ \mu\left(\pd B{y^{j}} - b_{j}\pd Bs \right) \;+\; \frac{p_{\|}^{2}}{m}\pd{b_{j}}s
\right], \label{eq:ya_dot} \\
\dot{s} & = & \frac{v_{\|}}{\mathcal{J}} \;-\; \frac{c\,\epsilon}{q\mathcal{J}}\;\eta^{ij}\left(\mu\;b_{i}\,\pd B{y^{j}} - \frac{p_{\|}^{2}}{m}
\pd{b_{i}}{y^{j}}\right), \label{eq:s_dot} \\
\dot{p}_{\|} & = & -\frac{\mu}{\mathcal{J}}\left[\pd Bs + \epsilon\frac{cp_{\|}}{q}\eta^{ij} \left(\pd{b_{i}}s\pd B{y^{j}} + \pd{b_{i}}{y^{j}}
\pd Bs\right)\right],
\label{eq:ppar_dot}
\end{eqnarray}
where the two-by-two antisymmetric tensor $\eta^{ij}$ has components
$\eta^{12}=-1=-\eta^{21}$, and the total guiding-center phase-space Jacobian $\mathcal{J}$, which includes the spatial Jacobian 
$(\nabla\alpha\btimes\nabla\beta\bdot\nabla s)^{-1}=B^{-1}$ and the guiding-center Jacobian $B_{\|}^{*}$, is 
\begin{equation}
\mathcal{J} \;\equiv\; \frac{B_{\|}^{*}}{B} \;=\; 1 \;+\; \epsilon\;\frac{cp_{\|}}{qB}\; \bhat\bdot\nabla\btimes\bhat \;=\; 1 \;-\; \epsilon\;
\frac{cp_{\|}}{q}\;\eta^{ij}\; \left(\pd{b_{j}}{y^{i}} \;-\; b_{i}\pd{b_{j}}s\right).
\label{eq:Jacobian_abs}
\end{equation}
Equation \eqref{eq:ya_dot} describes the slow guiding-center motion
across magnetic field lines (in the magnetic-field label-space). Equations
\eqref{eq:s_dot}-\eqref{eq:ppar_dot}, on the other hand, describe
the fast guiding-center parallel dynamics (at lowest order), with higher-order
corrections involving magnetic-field curvature and the non-orthogonality
of the magnetic coordinates. Note that the guiding-center equations
\eqref{eq:ya_dot}-\eqref{eq:ppar_dot} exactly satisfy the Liouville identity
\[ \pd{}{y^{i}}\left(\mathcal{J}\frac{}{}\dot{y}^{i}\right) \;+\; \pd{}{s}\left(\mathcal{J}\frac{}{}\dot{s}\right) \;+\; \pd{}{p_{\|}}\left(
\mathcal{J}\frac{}{}\dot{p}_{\|}\right) \;\equiv\; 0, \]
which ensures that the four-dimensional guiding-center dynamics conserves the guiding-center phase-space volume.

The lowest-order guiding-center equations motion are obtained by setting the ordering parameter $\epsilon=0$ in 
Eqs.~\eqref{eq:s_dot}-\eqref{eq:Jacobian_abs}, so that $\dot{y}_{0}^{i} \equiv 0$ and
\begin{equation}
\dot{s}_{0} \;\equiv\; v_{\|} \;\;\;\;{\rm and}\;\;\;\; \dot{p}_{\|0} \;\equiv\; \dot{s}_{0}\;\pd{p_{\|}}s \;=\; -\;\mu\pd Bs.
\label{eq:par_0}
\end{equation}
These equations describe the lowest-order guiding-center parallel 
dynamics, where the parallel momentum $p_{\|}$ is treated in Eq.~\eqref{eq:par_0}
as a function of the magnetic coordinates $(\alpha,\beta,s)$ as well
as the guiding-center invariants $(\mathcal{E},\mu)$: 
\begin{equation}
|p_{\|}|(\alpha,\beta,s;\mathcal{E},\mu) \;=\; \sqrt{2m\,\left[ \mathcal{E} - \mu B(\alpha,\beta,s)\right]}.
\label{eq:ppar_def}
\end{equation}
By conservation of energy $\mathcal{E}$ and magnetic moment $\mu$,
a particle following a guiding-center orbit may encounter a turning
point along a field line where $p_{\|}$ vanishes (at a bounce point
$s=s_{\mathrm{b}}$ where $\dot{s}_{0}\equiv0$) and a guiding-center
becomes \textit{trapped} between two such turning points $s_{\mathrm{b}}^{\pm}$.

The lowest-order guiding-center motion \eqref{eq:par_0} is periodic on the $(s,p_{\|})$-plane. The area enclosed by the
lowest-order periodic orbit can be used to define the bounce action
\begin{equation}
J(\alpha,\beta;\mathcal{E},\mu) \;\equiv\; \frac{1}{2\pi}\;\oint p_{\|}\, ds \;=\; \frac{1}{\pi} \int_{s_{\rm b}^{-}}^{s_{\rm b}^{+}}\;
|p_{\|}|(\alpha,\beta,s;\mathcal{E},\mu)\; ds,
\label{eq:J_bounce_def}
\end{equation}
and the canonically-conjugate bounce angle 
\begin{equation}
\zeta \;\equiv\; \pi \;+\; {\rm sgn}(p_{\|})\;\omega_{\mathrm{b}}\;\int_{s_{\mathrm{b}}^{-}}^{s}\frac{ds^{\prime}}{|v_{\|}|(s^{\prime})}.
\label{eq:zeta_bounce_def}
\end{equation}
In Eq.~\eqref{eq:zeta_bounce_def}, the bounce angle is defined
so that $\zeta=\pi$ at the turning point $s_{\mathrm{b}}^{-}$ and
the bounce frequency $\omega_{\mathrm{b}}\equiv2\pi/\tau_{\mathrm{b}}$
is defined in terms of the bounce-period integral 
\begin{equation}
\tau_{\mathrm{b}}(\alpha,\beta;\mathcal{E},\mu) \;\equiv\; \oint\frac{ds}{v_{\|}} \;=\; 2\,\int_{s_{\mathrm{b}}^{-}}^{s_{\mathrm{b}}^{+}}
\frac{ds}{|v_{\|}|}.
\label{eq:tau_b}
\end{equation}
We note that, according to the standard guiding-center time-scale ordering, the ratio $\omega_{\rm b}/|\omega_{\rm g}| \equiv {\cal O}(\epsilon)$ is small.

\subsection{Three-dimensional guiding-center dynamics}

We note that if we replace the guiding-center parallel momentum $p_{\|}$ with the guiding-center energy $\mathcal{E}$, by using the relation
\eqref{eq:ppar_def} and using the guiding-center energy conservation law $\dot{{\cal E}} \equiv 0$, we may reduce the description of guiding-center dynamics to three coordinates $(\alpha,\beta,s)$:
\begin{eqnarray}
\dot{y}^{i} & = & \eta^{ij}\frac{c\epsilon}{q\mathcal{J}^{\mathcal{E}}}\left[ -\;\pd{p_{\|}}{y^{j}} \;+\; \pd{}s \left( p_{\|}b_{j}\right)\right]
\;=\; \frac{c\epsilon}{q}\,\eta^{ij} \left[\; \mu\;\pd{B}{y^{j}} \;+\; v_{\|}\;\pd{}{s} \left( |p_{\|}|\frac{}{}b_{j}\right) \;\right],
\label{eq:ya_dot_E}\\
\dot{s} & = & \frac{1}{\mathcal{J}^{\mathcal{E}}}\;\left[ \frac{v_{\|}}{|v_{\|}|} \;-\; \frac{c\epsilon}{q}\;\eta^{ij}\pd{}{y^{i}}\left(p_{\|}b_{j}\right)\right] \;=\; v_{\|}\; \left[\; 1 \;-\; \frac{c\epsilon}{q}\;\eta^{ij}\,\pd{}{y^{i}} \left( |p_{\|}|\frac{}{}b_{j}\right) \;\right],
\label{eq:s_dot_E}
\end{eqnarray}
where the new total guiding-center phase-space Jacobian is $\mathcal{J}^{\mathcal{E}}\equiv\mathcal{J}/
|v_{\|}|$, so that the Liouville identity for guiding-center dynamics now simply reads
\[ \pd{}{y^{i}}\left(\mathcal{J}^{\mathcal{E}}\frac{}{}\dot{y}^{i}\right) \;+\; \pd{}{s}\left(\mathcal{J}^{\mathcal{E}}\frac{}{}\dot{s}\right) \;\equiv\; 0. \]
The guiding-center motion \eqref{eq:ya_dot_E}-\eqref{eq:s_dot_E} is thus represented in a three-dimensional phase space $(\alpha,\beta,s;
\mathcal{E},\mu)$, where the rapid bounce motion $s(\zeta, J)$ is parameterized by the bounce action-angle coordinates $(J,\zeta)$, while the slow drift motion takes place in the two-dimensional space of magnetic-field labels $(\alpha,\beta)$.

\subsection{Bounce-action adiabatic invariance}

While the bounce action \eqref{eq:J_bounce_def} is an invariant for
the lowest-order parallel guiding-center dynamics (at zeroth order
in $\epsilon$ when the magnetic-field labels $\alpha$ and $\beta$
are frozen), it is not conserved at higher order. Indeed, we find
\begin{equation}
\frac{dJ}{dt} \;=\; \dot{y}^{i}\pd J{y^{i}} \;=\; -\;\frac{\dot{y}^{i}}{2\pi}\oint\mu\pd B{y^{i}}\frac{ds}{v_{\|}}.
\label{eq:Jb_dot}
\end{equation}
By introducing the bounce-angle averaging procedure 
\begin{equation}
\langle\cdots\rangle_{\mathrm{b}} \;\equiv\; \sum_{\pm}\frac{1}{\tau_{\mathrm{b}}}\int_{s_{\mathrm{b}}^{-}}^{s_{\mathrm{b}}^{+}}(\cdots)\;
\frac{ds}{|v_{\|}|},
\label{eq:bounce_average}
\end{equation}
(where $\sum_{\pm}$ denotes a sum of the sign of $v_{\|}$) we note
that, using Eq.~\eqref{eq:ya_dot_E} with $\mathcal{J}=1$ (at lowest
order in $\epsilon$), we find 
\[ \left\langle \mu\pd B{y^{i}}\right\rangle _{\mathrm{b}} \;=\; \epsilon^{-1}\;\frac{q}{c}\,\eta_{ij}\left\langle \dot{y}^{j}\right\rangle_{\mathrm{b}}, \]
where the two-by-two antisymmetric tensor $\eta_{ij}$ has components
$\eta_{12}=1=-\eta_{21}$ (i.e., $\eta_{ij}\eta^{jk}\equiv\delta_{i}^{k}$).
Equation \eqref{eq:Jb_dot} therefore becomes 
\begin{equation}
\frac{dJ}{dt} \;=\; -\;\frac{\dot{y}^{i}}{\omega_{\mathrm{b}}}\left\langle \mu\pd B{y^{i}}\right\rangle_{\mathrm{b}} \;=\; -\;
\frac{q\epsilon^{-1}}{c\,\omega_{\mathrm{b}}} \left( \eta_{ij}\frac{}{}\dot{y}^{i}\left\langle \dot{y}^{j}\right\rangle_{\mathrm{b}} \right) \;=\; -\;
\frac{q\epsilon^{-1}}{c\,\omega_{\mathrm{b}}}\left(\dot{\alpha}\,\langle \dot{\beta}\rangle_{\rm b} \;-\frac{}{} \dot{\beta}\,\langle 
\dot{\alpha}\rangle_{\mathrm{b}}\right),
\label{eq:Jb_dot_final}
\end{equation}
which satisfies the standard condition $\langle dJ/dt\rangle_{\mathrm{b}}\equiv 0$
for adiabatic invariants \cite{Cary_Brizard}.

The general theory of adiabatic invariance \cite{Northrop} for charged-particle
motion in magnetic fields allows the construction a new adiabatic
invariant $\ov J$ that is preserved up to second order in $\epsilon$.
The resulting asymptotic expansion $\ov J=\ov J_{0}+\epsilon\ov J_{1}+\cdots$,
where the lowest-order term $\ov J_{0}\equiv J$ is given by Eq.~\eqref{eq:J_bounce_def},
satisfies the condition for adiabatic invariance: 
\[ 0 \;=\; \frac{d\ov J}{dt} \;=\; \epsilon\left(\frac{d_{1}J}{dt} \;+\; \frac{d_{0}\ov J_{1}}{dt}\right) \;+\; \cdots, \]
where we use the expansion for the guiding-center operator $d/dt\equiv d_{0}/dt+\epsilon d_{1}/dt+\cdots$,
with the lowest-order operator $d_{0}/dt\equiv\omega_{\mathrm{b}}\;\partial/\partial\zeta$
expressed in terms of the bounce-angle derivative. We can now see
that $d\ov J/dt= {\cal O}(\epsilon^{2})$ is satisfied at first order in $\epsilon$ if
the first-order correction $\ov J_{1}$ is given as \cite{Northrop}
\begin{equation}
\ov J_{1} \;=\; -\;\frac{1}{\omega_{\mathrm{b}}}\;\int\frac{d_{1}J}{dt}d\zeta \;=\; \frac{q\epsilon^{-2}}{c\,\omega_{\mathrm{b}}^{2}}\;
\int\left(\dot{\alpha}\,\langle \dot{\beta}\rangle_{\rm b} \;-\frac{}{} \dot{\beta}\,\langle \dot{\alpha}\rangle_{\mathrm{b}} \right) d\zeta.
\label{eq:Jb_1}
\end{equation}
In order to complete the transformation from guiding-center coordinates $(\alpha,\beta, s(J,\zeta),p_{\|}(J,\zeta))$ to bounce-center coordinates 
$(\ov{\alpha},\ov{\beta},\ov J,\ov{\zeta})$, we need to find first-order corrections to the remaining bounce-center coordinates $(\ov{\alpha},\ov{\beta},\ov{\zeta})$.

\subsection{Lie-transform analysis}

The bounce-center transformation \cite{RGL_82a,AJB_00}
\begin{equation}
z^{a} \;=\; (\alpha,\beta,J,\zeta) \;\rightarrow\; \ov z^{a} \;=\; (\ov{\alpha},\ov{\beta},\ov J,\ov{\zeta})
\label{eq:bct_def}
\end{equation}
is a near-identity transformation on the four-dimensional guiding-center phase space: 
\begin{equation}
\ov z^{a} \;=\; z^{a} \;+\; \epsilon\; G_{1}^{a} \;+\; \cdots,
\label{eq:zbar_a}
\end{equation}
which is generated by vector fields $({\sf G}_{1}, {\sf G}_{2}, \cdots)$ where, at $n^{\rm th}$-order, the component $G_{n}^{a}$ removes the fast-angle dependence in the dynamical equation for $d\ov{z}^{a}/dt$. Here, the component $G_{1}^{J}=\ov J_{1}$ is already given by Eq.~\eqref{eq:Jb_1}, while the remaining components $(G_{1}^{\alpha},G_{1}^{\beta},G_{1}^{\zeta})$ are determined by Lie-transform perturbation method.

As a result of the bounce-center transformation \eqref{eq:zbar_a},
the guiding-center phase-space Lagrangian \eqref{eq:psl_alphabeta}
is transformed into the new bounce-center phase-space Lagrangian 
\begin{equation}
\ov{\Gamma}_{\mathrm{b}} \;\equiv\; \mathsf{T}_{\mathrm{b}}^{-1}\Gamma_{\mathrm{g}} \;+\; \exd S \;=\; \epsilon^{-1}\Gamma_{\mathrm{g}0} \;+\;
\left(\Gamma_{\mathrm{g}1} \;-\; \pounds_{1}\Gamma_{\mathrm{g}0} \;+\; \exd S_{1}\right) + \cdots,
\label{q:psl_bc}
\end{equation}
where the left side is given by Eq.~\eqref{eq:psl_bc}. From the first-order $(J,\zeta)$-components of Eq.~\eqref{q:psl_bc}, we obtain the
relations for the gauge function $S_{1}(\alpha,\beta;J,\zeta)$: 
\begin{equation}
\left( \begin{array}{c}
0 \\
\\
J
\end{array} \right) \;=\; p_{\|}\; \left( \begin{array}{c}
\partial s/\partial J \\
\\
\partial s/\partial \zeta
\end{array} \right) \;+\; \left( \begin{array}{c}
\partial S_{1}/\partial J \\
\\
\partial S_{1}/\partial \zeta
\end{array} \right),
\label{eq:S1_Jzeta}
\end{equation}
while the components $G_{1}^{i}=(G_{1}^{\alpha},G_{1}^{\beta})$
are determined from the relation 
\begin{equation}
G_{1}^{i} \;=\; \frac{c}{q}\eta^{ij}\left(p_{\|}b_{j} \;+\; \pd{S_{1}}{y^{j}}\right).
\label{eq:G1_i}
\end{equation}
The bounce action-angle components $(G_{1}^{J},G_{1}^{\zeta})$, on the other hand, are generally obtained either at higher order in the Lie-perturbation analysis or are more directly obtained from the equations of motion. For example, since the bounce-center equations of motion $d\ov z^{\alpha}/dt$ are independent of the bounce angle, we must have
\begin{equation}
\frac{d\ov z^{a}}{dt} \;\equiv\; \left\langle \frac{dz^{a}}{dt}\right\rangle_{\rm b} \;=\; \frac{dz^{a}}{dt} \;+\; \epsilon\,\left(
\omega_{\mathrm{b}}\;\pd{G_{1}^{a}}{\zeta} \;+\; \cdots \right), 
\label{eq:ovz_dot}
\end{equation}
from which we obtain
\begin{equation}
G_{1}^{a} \;\equiv\; \langle G_{1}^{a}\rangle_{\rm b} \;+\; \frac{1}{\epsilon\,\omega_{\mathrm{b}}}\;\int\,\left(\left\langle \frac{dz^{a}}{dt}\right
\rangle_{\mathrm{b}} \;-\; \frac{dz^{a}}{dt} \right)\; d\zeta,
\label{eq:G1_a_def}
\end{equation}
where the bounce-angle-independent part $\langle G_{1}^{a}\rangle_{\rm b}$ is computed at higher order.

Since the bounce-center transformation \eqref{eq:zbar_a} is a canonical transformation, its Jacobian $\ov{{\cal J}}$ is equal to the guiding-center Jacobian ${\cal J}$. By using the general relation
\begin{equation}
\ov{{\cal J}} \;\equiv\; {\cal J} \;-\; \epsilon\;\pd{}{z^{a}} \left( {\cal J}\frac{}{}G_{1}^{a} \right) \;+\; \cdots \;\equiv\; {\cal J},
\label{eq:ov_Jac}
\end{equation}
we easily find that
\begin{equation}
\pd{G_{1}^{i}}{y^{i}} \;=\; \frac{c}{q}\;\eta^{ij}\;\pd{}{y^{i}}\left(p_{\|}\frac{}{}b_{j}\right) \;=\; \epsilon^{-1}{\cal J}^{\cal E} \left( 
v_{\|} \;-\frac{}{} \dot{s} \right) \;=\; - \left( \pd{G_{1}^{J}}{J} \;+\; \pd{G_{1}^{\zeta}}{\zeta} \right),
\label{eq:G1i_yi}
\end{equation}
where we used Eq.~\eqref{eq:s_dot_E}. 

We now make a few remarks concerning Eqs.~\eqref{eq:S1_Jzeta}-\eqref{eq:G1_i}. First, in order to solve for the components $G_{1}^{i}$, we need
to solve for the gauge function $S_{1}$, which is determined from the relations \eqref{eq:S1_Jzeta}. Second, we can
use these relations to obtain the identity 
\[ 1 \;-\; \pd{}J\left(p_{\|}\pd s{\zeta}\right) \;=\; \frac{\partial^{2}S_{1}}{\partial\zeta\partial J} \;=\; -\;\pd{}{\zeta}\left(p_{\|}
\pd sJ\right), \]
which establishes that the relation $(s,p_{\|})\rightarrow(J,\zeta)$ satisfies the canonical identity
\begin{equation}
\{ s,\; p_{\|}\}_{\|} \;\equiv\; \pd s{\zeta}\pd{p_{\|}}J \;-\; \pd sJ\pd{p_{\|}}{\zeta} \;\equiv\; 1.
\label{eq:canonical_id}
\end{equation}
Hence, the solution for $S_{1}$ requires explicit expressions for the parallel-dynamic coordinates $(s,p_{\|})$ as functions of bounce action-angle coordinates $(J,\zeta)$. Since these expressions are generally not known beyond the deeply-trapped approximation (in which the bounce motion is represented as a simple-harmonic oscillation between turning points), the bounce-center analysis can only proceed forward on a formal basis (beyond the deeply-trapped approximation) by taking the canonical identity \eqref{eq:canonical_id} as an axiom of bounce-center transformation theory \cite{RGL_82a}. We note that the equivalent canonical relation in guiding-center theory is expressed as 
\begin{equation}
\left\{ \vb{\rho}_{\rm g},\frac{}{} {\bf p}_{\bot}\right\}_{\bot} \;\equiv\; {\bf I} - \bhat\,\bhat, 
\label{eq:canonical_gc}
\end{equation}
where $\vb{\rho}_{\rm g}$ denotes the guiding-center gyroradius vector (which explicitly depends on the gyroangle $\zeta_{\rm g}$), the perpendicular momentum is ${\bf p}_{\bot} \equiv m\Omega\,\partial\vb{\rho}_{\rm g}/\partial\zeta_{\rm g}$, and the perpendicular Poisson bracket $\{\;,\;\}_{\bot}$ involves derivatives with respect to the gyro-action $J_{\rm g} \equiv (mc/q)\,\mu$ and the gyroangle $\zeta_{\rm g}$. The two-dimensional canonical relation \eqref{eq:canonical_gc} can be proved explicitly because the relations $\vb{\rho}_{\rm g}(J_{\rm g},\zeta_{\rm g})$ and 
${\bf p}_{\bot}(J_{\rm g},\zeta_{\rm g})$ are known.

\section{\label{sec:gc_dipole}Guiding-center Motion in Axisymmetric Dipole geometry}

The motivation for the present work is to explore the canonical relation $(s,p_{\|})\rightarrow(J,\zeta)$, and the associated canonical identity 
\eqref{eq:canonical_id}, in a magnetic geometry simple enough to allow the analysis of bounce motion beyond the deeply-trapped approximation. 

\subsection{Axisymmetric Dipole Geometry}

For this purpose, we consider a pure-dipole magnetic field 
\begin{equation}
\mathbf{B} \;=\; \frac{\mathcal{M}}{r^{3}}\left(2\cos\theta\;\wh r \;+\; \sin\theta\;\wh{\theta}\right),
\label{eq:Bdip_def}
\end{equation}
where spherical coordinates $\left(r,\theta,\phi\right)$ are used
and the constant $\mathcal{M}\equiv B_{\mathrm{e}}r_{\mathrm{e}}^{3}$
combines the strength of the magnetic field $B_{\mathrm{e}}$ at the
equatorial radial distance $r_{\mathrm{e}}$ (at $\theta=\pi/2$). The strength of the dipole magnetic field at a point $(r,\theta)$ in
the ``poloidal'' plane (at fixed toroidal angle $\phi$) is
\[ B(r,\theta) = B_{\rm e}\;\frac{r_{\rm e}^{3}}{r^{3}}\; \sqrt{1 + 3\,\cos^{2}\theta}, \] 
while the parallel unit vector is
\[ \bhat \;=\; \frac{2\cos\theta\;\wh{r} \;+\; \sin\theta\;\wh{\theta}}{\sqrt{1 + 3\,\cos^{2}\theta}}. \]
We note that, in the axisymmetric dipole magnetic geometry represented by Eq.~\eqref{eq:Bdip_def}, the toroidal angle $\phi$ is an ignorable coordinate

We can also write the magnetic field \eqref{eq:Bdip_def} in 2-covariant
flux-coordinate representation 
\begin{equation}
\mathbf{B} \;=\; \nabla\psi\btimes\nabla\phi,
\label{eq:B_2cov}
\end{equation}
where the poloidal flux $\psi$ is 
\begin{equation}
\psi\left(r,\theta\right) \;=\; \psi_{\rm e}\;\frac{r_{\rm e}}{r}\sin^{2}\theta.
\label{eq:psi_def}
\end{equation}
Since a magnetic field line lies on a constant-$\psi$ surface $\psi = \psi_{\rm e}$, we find that 
\begin{equation}
r\left(r_{\mathrm{e}},\theta\right) \;=\; r_{\mathrm{e}}\sin^{2}\theta
\label{eq:r_theta}
\end{equation}
draws a single magnetic field line on the magnetic surface $\psi = \psi_{\rm e}$. This relation yields the distance
along a single magnetic-field line
\begin{widetext}
\begin{eqnarray}
s\left(r_{\mathrm{e}},\theta\right) & \equiv & \frac{r\left(r_{\mathrm{e}},\theta\right)}{\sin^{2}\theta}\;\int_{\pi/2}^{\theta}
\sqrt{1+3\cos^{2}\theta^{\prime}}\sin\theta^{\prime}d\theta^{\prime} \;=\; r_{\mathrm{e}}\;\int_{\pi/2}^{\theta}\sqrt{1+3\cos^{2}\theta^{\prime}}\sin\theta^{\prime}d\theta^{\prime} \nonumber \\
 & = & \frac{r_{\mathrm{e}}}{2\sqrt{3}}\left[ \ln\left( \sqrt{1+3\cos^{2}\theta}-\sqrt{3}\cos\theta\right) \;-\;
\sqrt{3}\cos\theta\sqrt{1+3\cos^{2}\theta}\;\right],
\label{eq:s_dipole}
\end{eqnarray}
\end{widetext}
which is positive for $\theta>\pi/2$ (below the equator), negative for $\theta<\pi/2$ (above the equator), and zero at the equator $\left(\theta=\pi/2\right)$. The length of a magnetic-field line on the constant-$\psi$ surface labeled by $r_{\mathrm{e}}$ is 
\begin{equation}
L_{\mathrm{e}} \;=\; r_{\mathrm{e}}\,\left[ 2 \;+\; \frac{1}{\sqrt{3}}\;\ln\left(2+\sqrt{3}\right)\right] \;=\; (2.76...)\;r_{\rm e}. 
\label{eq:Le_def}
\end{equation}
The magnetic-field strength on a single line (on a constant-$\psi$ surface labeled by $r_{\mathrm{e}}$), on the other hand, is given as 
\begin{equation}
B(\theta) \;=\; B_{\mathrm{e}}\; \frac{\sqrt{1+3\cos^{2}\theta}}{\sin^{6}\theta},
\label{eq:B_dipole}
\end{equation}
which implies that all particles are trapped in this pure-dipole field (since $B$ becomes infinite as $\theta\rightarrow0$ or $\pi$).

The parallel unit vector $\bhat$ associated with Eq.~\eqref{eq:B_2cov}
is expressed in terms of the coordinates $\left(\psi,\phi,s\right)$ as 
\begin{equation}
\bhat \;=\; \nabla s \;+\; a\left(\psi,s\right)\nabla\psi,
\label{eq:bhat_psis}
\end{equation}
where the covariant components $(b_{\psi}, b_{\phi}, b_{s})$ are
\[ b_{\psi} \;\equiv\; \bhat\bdot\pd{\mathbf{X}}{\psi} \;=\; a\left(\psi,s\right) \]
$b_{\phi}\equiv\bhat\bdot\partial\mathbf{X}/\partial\phi=0$ and $b_{s} = 1$. We note that, since $a \equiv -\,\nabla s\bdot\nabla\psi/|\nabla
\psi|^{2}$, the magnetic coordinates $(\psi,\phi,s)$ are non-orthogonal, i.e.,
\begin{equation}
a(r,\theta) \;=\; \frac{r}{\psi} \left( \frac{s}{r} \;-\; \frac{2\,\cos\theta}{\sqrt{1 + 3\,\cos^{2}\theta}} \right) \;\neq\; 0,
\label{eq:a_def}
\end{equation}
off the equatorial plane $(\theta \neq \pi/2$). Moreover, the magnetic curvature $\bhat\bdot\nabla\bhat = \nabla\psi\;\partial a/\partial s$ is purely perpendicular to the magnetic surface.

\subsection{Guiding-center motion in axisymmetric dipole geometry}

In axisymmetric dipole geometry, the total guiding-center Jacobian \eqref{eq:Jacobian_abs} is simply $\mathcal{J} = 1$ (since 
$\bhat\bdot\nabla\btimes\bhat = 0$), with ${\cal J}^{{\cal E}} = 1/|v_{\|}|$, and the guiding-center drift equations of motion \eqref{eq:ya_dot_E} become 
\begin{eqnarray}
\dot{\psi} & = & 0, \label{eq:psidot_dip}\\
\dot{\phi} & = & \frac{c\epsilon}{q}\left[\mu\pd B{\psi}+v_{\|}\pd{}s\left(p_{\|}a\right)\right],
\label{eq:phidot_dip}
\end{eqnarray}
where Eq.~\eqref{eq:psidot_dip} follows from the axisymmetry of the magnetic-dipole field. A trapped-particle orbit is characterized by a pitch-angle
coordinate $\xi\equiv p_{\|}/p$ which vanishes at the turning points $\theta_{\mathrm{b}}$. Conservation of magnetic moment and energy yields the relation 
\begin{equation}
\frac{\mu}{\mathcal{E}} \;=\; \frac{1-\xi^{2}\left(\theta\right)}{B\left(\theta\right)} \;=\; \frac{1-\xi_{\mathrm{e}}^{2}}{B_{\mathrm{e}}},
\label{eq:xi_theta}
\end{equation}
and hence, for a fixed value $\xi_{\mathrm{e}}$, the turning-point angle $\theta_{\mathrm{b}}$ is a solution of the relation 
\begin{equation}
\frac{B\left(\theta_{\mathrm{b}}\right)}{B_{\mathrm{e}}} \;=\; \left( 1 \;-\; \xi_{\mathrm{e}}^{2}\right)^{-1},
\label{eq:Bxi_turning}
\end{equation}
since $\xi(\theta_{\rm b}) \equiv 0$. By using Eqs.~\eqref{eq:r_theta} and \eqref{eq:B_dipole}, the toroidal drift precession angular frequency \eqref{eq:phidot_dip} for a guiding-center with invariants $(\psi,{\cal E},\mu)$ can be expressed as
\begin{widetext}
\begin{equation}
\dot{\phi}_{\rm e}(\theta) \;=\; \frac{3c\,{\cal E}}{q\,B_{\rm e}\,r_{\rm e}^{2}}\;
\frac{\sin^{2}\theta\,(1 + \cos^{2}\theta)}{(1 + 3\,\cos^{2}\theta)^{2}}
\left[\; 2 \;-\; \left(1 - \xi_{\rm e}^{2}\right)\; \frac{\sqrt{1 + 3\,\cos^{2}\theta}}{\sin^{6}\theta} \;\right],
\label{eq:phi_theta}
\end{equation}
\end{widetext}
where $\xi_{\rm e} \equiv p_{\|{\rm e}}/p = \sqrt{1 - \mu\,B_{\rm e}/{\cal E}}$ defines the pitch-angle coordinate at the equator. 

The guiding-center parallel equations of motion \eqref{eq:s_dot}-\eqref{eq:ppar_dot}, on the other hand, become
\begin{eqnarray}
\dot{s} & = & v_{\|},\label{eq:sdot_dip}\\
\dot{p}_{\|} & = & -\;\mu\;\pd Bs,
\label{eq:p_pardot_dip}
\end{eqnarray}
which immediately imply that
the bounce action \eqref{eq:J_bounce_def} is exactly conserved in
axisymmetric dipole geometry since 
\begin{equation}
J(\psi;\mathcal{E},\mu) \;=\; \frac{1}{2\pi}\oint p_{\|}ds
\label{eq:J_dipole}
\end{equation}
is now explicitly a function of the guiding-center invariants. Lastly, we note that, by Noether's Theorem, the drift action in axisymmetric dipole geometry is simply $J_{\mathrm{d}}\equiv q\psi/c$.

\subsection{Bounce action-angle coordinates}

\begin{figure}
\epsfysize=2.5in
\epsfbox{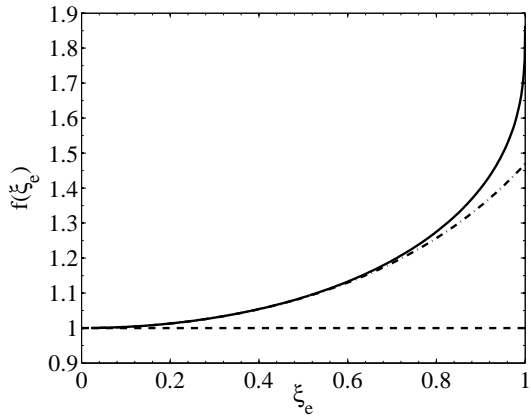}
\caption{Plot of the normalized bounce-period function $f\left(\xi_{\mathrm{e}}\right)$
as a function of the equatorial pitch-angle coordinate $\xi_{\mathrm{e}}$
(solid line). In the deeply-trapped approximation (dashed line), $f\left(\xi_{\mathrm{e}}\right)$
is replaced by unity in Eq.~\eqref{eq:tau_b_xi}. The dash-dotted line includes the first-order (anharmonic) correction to be discussed in 
Sec.~\ref{subsec:anharm}.}
\label{fig:f_xi} 
\end{figure}

A trapped-particle orbit can be represented in terms of the canonical action-angle coordinates \eqref{eq:J_bounce_def}-\eqref{eq:zeta_bounce_def},
where the bounce action is 
\begin{equation}
J(r_{\mathrm{e}},p,\xi_{\mathrm{e}}) \;=\; \frac{r_{\mathrm{e}}p}{2\pi}\oint
|\xi|\left(\theta,\xi_{\mathrm{e}}\right)\, \sqrt{1+3\cos^{2}\theta}\sin\theta\,d\theta
\label{eq:J_def}
\end{equation}
and the conjugate bounce angle is 
\begin{equation}
\zeta \;=\; \pi \;\pm\; \frac{r_{\mathrm{e}}\,\omega_{\mathrm{b}}}{p/m}\int_{\theta_{\mathrm{b}}^{-}}^{\theta}
\frac{\sqrt{1+3\cos^{2}\theta^{\prime}}\sin\theta^{\prime}}{|\xi|\left(\theta^{\prime},\xi_{\mathrm{e}}\right)}\;d\theta^{\prime},
\label{eq:zeta_def}
\end{equation}
where $\pm={\rm sgn}(p_{\|})$ and the bounce frequency $\omega_{\mathrm{b}}$ is defined as 
\begin{equation}
\omega_{\mathrm{b}}^{-1} \;\equiv\; \pd J{\mathcal{E}} \;=\; \frac{mr_{\mathrm{e}}}{2\pi p}\oint\;
\frac{\sqrt{1+3\cos^{2}\theta}\sin\theta}{|\xi|\left(\theta,\xi_{\mathrm{e}}\right)}\;d\theta.
\label{eq:omega_b}
\end{equation}
In Eqs.~\eqref{eq:J_def}-\eqref{eq:omega_b}, the magnitude of the pitch-angle coordinate
\begin{equation}
|\xi|\left(\theta,\xi_{\mathrm{e}}\right) \;=\; \sqrt{1 \;-\; (1 - \xi_{\rm e}^{2})\;\frac{\sqrt{1+3\cos^{2}\theta}}{\sin^{6}\theta}}
\label{eq:xi_theta_e}
\end{equation}
is a function of the equatorial pitch-angle coordinate $\xi_{\rm e}$ and vanishes at the bounce points $\theta_{\rm b}^{\pm}$. When the integral in 
Eq.~\eqref{eq:omega_b} is computed numerically as a function of $\xi_{\mathrm{e}}$, we obtain the following expression for the bounce period \cite{Parks}
\begin{equation}
\tau_{\mathrm{b}} \;=\; \frac{2\pi}{\omega_{\mathrm{b}}} \;\equiv\; \frac{2\pi}{\Omega_{\mathrm{b}}}\; f\left(\xi_{\mathrm{e}}\right) \;\simeq\;
\frac{2\pi}{\Omega_{\mathrm{b}}}\left(1 \;+\; \frac{23}{72}\;\xi_{\mathrm{e}}^{2} \right),
\label{eq:tau_b_xi}
\end{equation}
where 
\begin{equation}
\Omega_{\mathrm{b}} \;=\; \frac{3\, v}{\sqrt{2}\, r_{\mathrm{e}}} \;=\; \left(\frac{3\,\rho_{\mathrm{e}}}{\sqrt{2}\, r_{\mathrm{e}}}\right)
|\omega_{{\rm ge}}|
\label{eq:Omega_b_def}
\end{equation}
denotes the bounce frequency of deeply-trapped particles (with $\xi_{\rm e} = 0$ and $\rho_{\mathrm{e}}\equiv v/|\omega_{{\rm ge}}|$) and the normalized function $f\left(\xi_{\mathrm{e}}\right)$ is shown in Fig.~\ref{fig:f_xi}. Note that, at $\xi_{\mathrm{e}}=1$ (for marginally-trapped particles), the bounce period \eqref{eq:tau_b_xi} is finite, with $f(1)=\sqrt{18}\, L_{\mathrm{e}}/(2\pi\, r_{\mathrm{e}})\simeq1.86$. Figure \ref{fig:f_xi} also shows the approximate expression for $f(\xi_{\rm e}) \simeq 1 + 23\,\xi_{\rm e}^{2}/72$ (dash-dotted line), which will be recovered from Lie-transform perturbation theory (see Sec.~\ref{subsec:anharm}).

\begin{figure}
\epsfysize=2.5in
\epsfbox{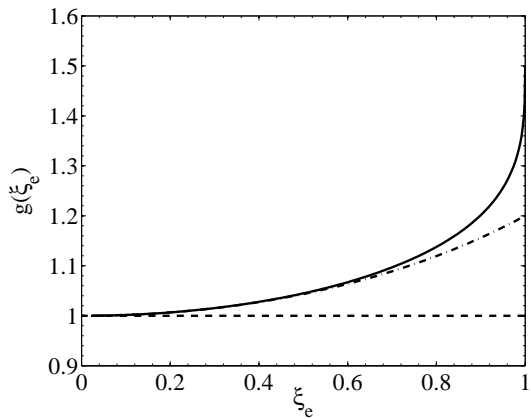}
\caption{Plot of the normalized drift-period function $g\left(\xi_{\mathrm{e}}\right)$
as a function of the equatorial pitch-angle coordinate $\xi_{\mathrm{e}}$
(solid line). In the deeply-trapped approximation (dashed line), $g\left(\xi_{\mathrm{e}}\right)$
is replaced by unity in Eq.~\eqref{eq:tau_d_xi}. The dash-dotted line includes the harmonic and anharmonic corrections discussed in Sec.~\ref{subsec:anharm}.}
\label{fig:g_xi} 
\end{figure}

Next, using Eq.~\eqref{eq:phi_theta}, the bounce-averaged drift frequency is obtained from Eq.~\eqref{eq:phi_theta} and is expressed as \cite{Hamlin}
\begin{widetext}
\begin{equation}
\omega_{\rm d} \;=\; \langle\dot{\phi}_{\rm e}\rangle_{\rm b} \;\equiv\; \frac{3\,\rho_{\rm e}}{2\tau_{\rm b}\,r_{\rm e}}\; 
\oint \frac{(1 + \cos^{2}\theta)\;d\theta}{(1 + 3\,\cos^{2}\theta)^{3/2}}\; \left[\; \frac{2\;\sin^{6}\theta \;-\; \left(1 - \xi_{\rm e}^{2} \right)\;
\sqrt{1 + 3\,\cos^{2}\theta}}{\sqrt{\sin^{6}\theta \;-\; \left(1 - \xi_{\rm e}^{2} \right)\;\sqrt{1 + 3\,\cos^{2}\theta}}} \;\right].
\label{eq:omega_d}
\end{equation}
\end{widetext}
where the bounce period $\tau_{\rm b}$ is defined in Eq.~\eqref{eq:tau_b_xi}. The drift period is therefore defined as
\begin{equation}
\tau_{\mathrm{d}} \;=\; \frac{2\pi}{\omega_{\mathrm{d}}} \;\equiv\; \frac{2\pi}{\Omega_{\mathrm{d}}}\; g\left(\xi_{\mathrm{e}}\right) \;\simeq\;
\frac{2\pi}{\Omega_{\mathrm{d}}} \left(1 \;+\; \frac{1}{6}\;\xi_{\rm e}^{2} \right),
\label{eq:tau_d_xi}
\end{equation}
where
\begin{equation}
\Omega_{\mathrm{d}} \;=\; \frac{3\,\rho_{\mathrm{e}}\, v}{2\, r_{\mathrm{e}}^{2}} \;=\; \left(\frac{\rho_{\mathrm{e}}}{\sqrt{2}\, r_{\mathrm{e}}}\right)\Omega_{\mathrm{b}}
\label{eq:Omega_d_def}
\end{equation}
denotes the drift frequency of deeply-trapped particles (with $\xi_{\rm e} = 0$) and the normalized function $g\left(\xi_{\mathrm{e}}\right)$ is shown in Fig.~\ref{fig:g_xi}, with $g(1) = 3/2$. Figure \ref{fig:g_xi} also shows the approximate expression $g(\xi_{\rm e}) \simeq 1 + \xi_{\rm e}^{2}
/6$ (dash-dotted line), which will be recovered from Lie-transform perturbation theory (see Sec.~\ref{subsec:anharm}). Using Eq.~\eqref{eq:phidot_dip}, we note that the bounce-averaged drift frequency \eqref{eq:omega_d} is also defined as
\begin{equation}
\omega_{\mathrm{d}} \;=\; \frac{c\mu}{q}\left\langle \pd B{\psi}\right\rangle_{\mathrm{b}}.
\label{eq:omega_d_xi}
\end{equation}
For deeply-trapped particles $(\xi_{\mathrm{e}}\ll1)$, Eq.~\eqref{eq:omega_d_xi} is approximately given as 
\begin{equation}
\Omega_{\rm d} \;=\; \lim_{\xi_{\mathrm{e}}\rightarrow0}\;\omega_{\rm d} \;=\; \frac{c\,{\cal E}}{qB_{\rm e}}\;\pd{B_{\mathrm{e}}}{\psi_{\mathrm{e}}} \;=\; \frac{3\;c\,{\cal E}}{q\,\psi_{\rm e}},
\label{eq:phi_dot_ave}
\end{equation}
from which we recover Eq.~\eqref{eq:Omega_d_def} in the deeply-trapped
limit (where $g\rightarrow1$). 

Lastly, when the bounce period \eqref{eq:tau_b_xi} and the gyroperiod $\tau_{\rm g} \equiv 2\pi/|\omega_{\rm g}|$ are compared to the bounce-averaged drift period \eqref{eq:tau_d_xi}, we obtain three well-separated orbital time scales $\tau_{\mathrm{g}}\ll \tau_{\mathrm{b}} \ll \tau_{\mathrm{d}}$ when $\rho_{\mathrm{e}}/r_{\mathrm{e}}\ll1$ for all values of $\xi_{\mathrm{e}}$ (since $1 \leq f, g < 2$).

\section{\label{sec:bc_deep}Bounce action-angle coordinates for deeply-trapped particles}

While the bounce action \eqref{eq:J_dipole} is now an exact invariant for guiding-center motion in an axisymmetric dipole magnetic field, we are still committed to investigating the canonical relation $(s,p_{\|})\rightarrow(J,\zeta)$ in this simple geometry. For this purpose, we first consider the lowest-order {\it harmonic} trapped-particle motion in the vicinity of the equator $\left(\theta\simeq\pi/2\right)$ and, then, consider the first-order 
{\it anharmonic} corrections.

We begin by introducing the latitude angle $\sigma\equiv\theta-\pi/2$ (with $|\sigma|\ll\pi/2$), so that the distance along a field 
line \eqref{eq:s_dipole} becomes 
\begin{widetext}
\begin{equation}
s \;=\; r_{\mathrm{e}}\left(\sin\sigma \;+\; \frac{1}{2}\sin^{3}\sigma \;+\; \cdots\right) \;=\; r_{\mathrm{e}}\left(\sigma \;+\; \frac{1}{3}\sigma^{3}
\;+\; \cdots\right),
\label{eq:s_sigma}
\end{equation}
\end{widetext}
i.e., in the vicinity of the equator, $|\sigma|=|s|/r_{\mathrm{e}}\ll 1$ represents a normalized parallel distance from the equatorial plane. In this approximation, the magnetic-field strength \eqref{eq:B_dipole} along a single field line yields the expression
\begin{equation}
\frac{B}{B_{\mathrm{e}}} \;=\; 1 \;+\; \frac{9}{2}\sin^{2}\sigma \;+\; \frac{75}{8}\sin^{4}\sigma \;+\; \cdots \;=\; 1 \;+\; \frac{9}{2}\sigma^{2} \;+\; \frac{63}{8}\sigma^{4} \;+\; \cdots,
\label{eq:B_sigma}
\end{equation}
while the unit covariant component (\ref{eq:a_def}) is expressed as
\begin{equation}
a \;=\; 3\;\frac{r_{\rm e}}{\psi_{\rm e}}\;\sigma \;+\; \cdots.
\label{eq:a_deep}
\end{equation}
Next, the particle energy reads 
\begin{equation}
\mathcal{E} \;=\; \frac{p_{\|}^{2}}{2m} \;+\; \mu B_{\mathrm{e}}\left( 1 \;+\; \frac{9}{2}\sigma^{2} \;+\; \cdots\right),
\label{eq:E_sigma}
\end{equation}
and hence, close to the equator, the guiding-center moves in the potential
$(9/2)\;\mu B_{\mathrm{e}}\;\sigma^{2}$ of a harmonic oscillator with
energy $\mathcal{E}-\mu B_{\mathrm{e}}$. (We show below that the
term $\mu B_{\mathrm{e}}$ acts as the potential for the bounce-averaged
toroidal drift motion.)

\begin{figure}
\epsfysize=2.5in
\epsfbox{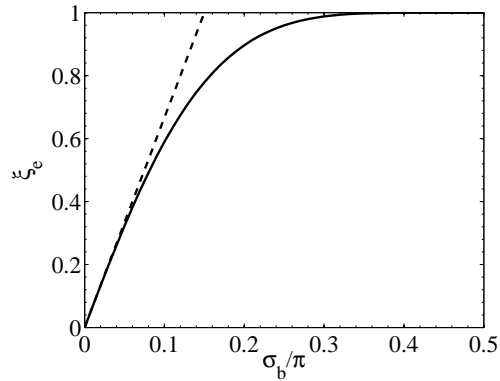}
\caption{Equatorial-pitch coordinate $|\xi_{\mathrm{e}}|$ versus the normalized bounce-latitude $\sigma_{\mathrm{b}}/\pi$.
The dashed line represents the linear approximation $|\xi_{\mathrm{e}}| \simeq 3\sigma_{\mathrm{b}}/\sqrt{2}$.}
\label{fig:eq_pitch} 
\end{figure}

Deeply-trapped particles have turning points located close to the equator (i.e., $|\xi_{\mathrm{e}}|\ll1$), so that we find $\theta_{\mathrm{b}}^{\pm}=\pi/2\pm\sigma_{\mathrm{b}}$,
and the equatorial pitch-angle coordinate is 
\begin{equation}
|\xi_{\mathrm{e}}| \;=\; \frac{3\,\sigma_{\mathrm{b}}}{\sqrt{2}} \left( 1 \;-\; \frac{11}{8}\sigma_{\mathrm{b}}^{2} \;+\; \cdots\right).
\label{eq:sigma_b}
\end{equation}
Figure \ref{fig:eq_pitch} shows the equatorial pitch-angle $|\xi_{\mathrm{e}}|$
as a function of the bounce latitude $\sigma_{\mathrm{b}}$ obtained from Eq.~\eqref{eq:Bxi_turning},
where we clearly see that, for deeply-trapped particles $\left(\sigma_{\mathrm{b}}\ll\pi/2\right)$,
the equatorial pitch-angle $|\xi_{\mathrm{e}}|$ depends linearly on $\sigma_{\mathrm{b}}$ (for up to $\xi_{\rm e} \simeq 0.6$).

\subsection{\label{subsec:lowest}Lowest-order harmonic calculations}

We now calculate explicit expressions for the bounce angle-action
coordinates \eqref{eq:J_def}-\eqref{eq:zeta_def} in the deeply-trapped approximation. First, in the vicinity of
the equator $(|\sigma| \ll \pi/2$), Eq.~\eqref{eq:xi_theta_e} yields
\begin{equation}
\left|\xi\right|\left(\theta,\xi_{\mathrm{e}}\right) \;=\; \frac{3}{\sqrt{2}}\;\sqrt{\sigma_{\mathrm{b}}^{2} - \sigma^{2}} \;+\; \cdots,
\label{eq:xi_deep}
\end{equation}
where terms of first order in the ordering parameter 
\begin{equation}
\epsilon_{\mathrm{t}} \;\equiv\; \sigma_{\mathrm{b}}^{2} \;=\; \left(\frac{s_{\mathrm{b}}}{r_{\mathrm{e}}}\right)^{2} \;\ll\; 1
\end{equation}
have been omitted beyond the {}``harmonic-oscillator'' approximation in Eq.~\eqref{eq:xi_deep}. The bounce action \eqref{eq:J_def} yields the lowest-order expression 
\begin{equation}
J_{0} \;=\; \frac{3r_{\mathrm{e}}p}{\pi\sqrt{2}}\int_{-\sigma_{\mathrm{b}}}^{\sigma_{\mathrm{b}}}\sqrt{\sigma_{b}^{2} - \sigma^{2}}\;d\sigma \;=\; 
\frac{3r_{\mathrm{e}}p}{2\sqrt{2}}\;\sigma_{\mathrm{b}}^{2} \;=\; \frac{{\cal E}\,\xi_{\mathrm{e}}^{2}}{\Omega_{\rm b}}.
\label{eq:J_deep}
\end{equation}
From this expression, we use Eq.~\eqref{eq:omega_b} to compute
the lowest-order bounce frequency 
\begin{equation}
\omega_{\mathrm{b}0} \;\equiv\; \left(\pd{J_{0}}{\mathcal{E}}\right)^{-1} \;=\; \Omega_{\mathrm{b}}\;\left( 1 \;-\; \epsilon_{\mathrm{t}}\;
\frac{\xi_{\mathrm{e}}^{2}}{2}\right)^{-1},
\label{eq:omegab_deep}
\end{equation}
where we used 
\begin{equation}
\pd{(p\,\xi_{\mathrm{e}}^{2})}{\mathcal{E}} \;=\; \frac{2\, m}{p}\; \left( 1 \;-\; \epsilon_{\mathrm{t}}\;\frac{\xi_{\mathrm{e}}^{2}}{2}\right).
\label{eq:pxi_E}
\end{equation}
When we combine these two expressions, where we henceforth use $\omega_{{\rm b}0} \simeq \Omega_{\rm b}$ to lowest order in $\epsilon_{\rm t}$ (the first-order correction will be used in Sec.~\ref{subsec:anharm}), we find that bounce harmonic-oscillator energy 
\begin{equation}
J_{0}\Omega_{\mathrm{b}} \;\equiv\; \frac{p^{2}\xi_{\mathrm{e}}^{2}}{2m} \;=\; \frac{p_{\|\mathrm{e}}^{2}}{2m} \;=\; \mathcal{E} - \mu B_{\mathrm{e}}\label{eq:JOmega}
\end{equation}
is simply the relative energy as measured from the minimum-$B$ potential
$\mu B_{\mathrm{e}}$. The lowest-order Hamiltonian therefore reads
\begin{equation}
H_{0} \;=\; \mu B_{\mathrm{e}} \;+\; J_{0}\,\Omega_{\mathrm{b}},
\label{eq:H0}
\end{equation}
and the lowest-order normalized bounce period $\Omega_{\rm b}\,(\partial H_{0}/\partial J_{0})^{-1} = 1$ is shown in Fig.~\ref{fig:f_xi} as a dashed line.

Next, we compute the lowest-order expression for the bounce angle $\zeta_{0}$ from Eq.~\eqref{eq:zeta_def}. First, using Eq.~\eqref{eq:xi_deep},
we compute 
\begin{widetext}
\begin{equation}
\int_{\theta_{\mathrm{b}}^{-}}^{\theta}\frac{\sqrt{1+3\cos^{2}\theta^{\prime}}\sin\theta^{\prime}}{|\xi|\left(\theta^{\prime},\xi_{\mathrm{e}}\right)}
\;d\theta^{\prime} \;\simeq\; \frac{\sqrt{2}}{3}\int_{-\sigma_{\mathrm{b}}}^{\sigma}\frac{d\sigma^{\prime}}{\sqrt{\sigma_{\mathrm{b}}^{2} - 
\sigma^{\prime2}}} \;=\; \frac{\sqrt{2}}{3}\left[\pi-\arccos\left(\frac{\sigma}{\sigma_{\mathrm{b}}}\right)\right],
\end{equation}
\end{widetext}
and thus we find the lowest-order bounce angle
\begin{equation}
\zeta_{0} \;=\;
\begin{cases}
2\pi-\arccos\left(\sigma/\sigma_{\mathrm{b}}\right) & \left(p_{\|}>0\right)\\
\arccos\left(\sigma/\sigma_{\mathrm{b}}\right) & \left(p_{\|}<0\right).
\end{cases}
\label{eq:zeta_deep}
\end{equation}
Lastly, the parallel momentum $p_{\|}$ and the parallel field-line coordinate $s$ are thus expressed (to lowest order in $\epsilon_{\rm t}$)
in terms of the action-angle coordinates $\left(J_{0},\zeta_{0}\right)$ as 
\begin{equation}
p_{\|} \;=\; -\;\left(2m\frac{}{}J_{0}\Omega_{\mathrm{b}}\right)^{1/2}\;\sin\zeta_{0} \;\equiv\; p_{\|0}(J_{0},\zeta_{0}),
\label{eq:ppar_deep}
\end{equation}
and 
\begin{equation}
s \;=\; \left(\frac{2J_{0}}{m\Omega_{\mathrm{b}}}\right)^{1/2}\;\cos\zeta_{0} \;\equiv\; s_{0}(J_{0},\zeta_{0}),
\label{eq:s_deep}
\end{equation}
where we have used the bounce frequency \eqref{eq:omegab_deep} for deeply-trapped particles. Equations \eqref{eq:ppar_deep}-\eqref{eq:s_deep} represent the standard canonical action-angle transformation for a harmonic oscillator and, thus, the lowest-order transformation
$\left(p_{\|0},s_{0}\right)\rightarrow\left(J_{0},\zeta_{0}\right)$ satisfies the canonical condition \eqref{eq:canonical_id}:
\begin{equation}
\{s_{0},p_{\|0}\}_{\|0} \;=\; \pd{s_{0}}{\zeta_{0}}\,\pd{p_{\|0}}{J_{0}} \;-\; \pd{s_{0}}{J_{0}}\,\pd{p_{\|0}}{\zeta_{0}} \;=\; \cos^{2}\zeta_{0} 
\;+\; \sin^{2}\zeta_{0} \;=\; 1,
\label{eq:PB_0}
\end{equation}
where partial derivatives are evaluated at constant ${\cal E}$ and $\psi_{\rm e}$. We note that deeply-trapped particles in axisymmetric tokamak geometry 
\cite{TSH_2009} are also described by the harmonic-oscillator representation \eqref{eq:ppar_deep}-\eqref{eq:s_deep}.

\subsection{\label{subsec:anharm}First-order anharmonic corrections}

We now proceed to obtain the first-order (anharmonic) corrections $(J_{1},\zeta_{1})$ for the bounce action-angle coordinates $(J=J_{0}+
\epsilon_{\mathrm{t}}J_{1},\zeta=\zeta_{0}+\epsilon_{\mathrm{t}}\zeta_{1})$ through the Hamiltonian Lie-transform perturbation method. By inverting 
the relation \eqref{eq:s_sigma} for $s(\sigma)$, we obtain 
\begin{equation}
\sigma(s) \;=\; \frac{s}{r_{\mathrm{e}}} \;-\; \frac{\epsilon_{\mathrm{t}}}{3}\;\left(\frac{s}{r_{\mathrm{e}}}\right)^{3} \;+\; \cdots,
\label{eq:sigma_s}
\end{equation}
which can be inserted in Eq.~\eqref{eq:B_sigma} to obtain 
\begin{equation}
\mu B \;=\; \mu B_{\mathrm{e}}\left[ 1 \;+\; \frac{9}{2}\left(\frac{s}{r_{\mathrm{e}}}\right)^{2} \;+\; \frac{39}{8}\left(\frac{s}{r_{\mathrm{e}}}\right)^{4} \;+\; \cdots\right].
\label{eq:muBs}
\end{equation}
The guiding-center Hamiltonian \eqref{eq:Ham_gc}, written as
\begin{equation}
H \;=\; H_{0}(J_{0}) \;+\; \epsilon_{\mathrm{t}}\; H_{1}\left(J_{0},\zeta_{0}\right),
\end{equation} 
will therefore have a lowest-order expression $H_{0}(J_{0})$ and a higher-order anharmonic correction $H_{1}(J_{0},\zeta_{0})$ that will
break the bounce-action invariance (i.e., $\dot{J_{0}} = -\,\epsilon_{\mathrm{t}}\;\partial H_{1}/\partial\zeta_{0}\neq0$). 
We will henceforth omit contributions of order higher than $\epsilon_{\mathrm{t}}$ beyond the harmonic-oscillator dynamics described in 
Sec.~\ref{subsec:lowest}.

Using the Lie-transform perturbation method, we perform a transformation to new bounce action-angle coordinates $\left(J,\zeta\right)$ in order to restore the invariance of the bounce-action, i.e., $\dot{J} = {\cal O}(\epsilon_{\rm t}^{2})$. Using the lowest-order bounce frequency \eqref{eq:omegab_deep}, the expression \eqref{eq:muBs} becomes 
\[ \mu B \;=\; \mu B_{\mathrm{e}} \;+\; \frac{m}{2}\Omega_{\mathrm{b}}^{2}s^{2}\left[1 \;-\; \epsilon_{\mathrm{t}}\;\xi_{\mathrm{e}}^{2} \;+\;
\epsilon_{\mathrm{t}}\frac{13}{12}\left(\frac{s}{r_{\mathrm{e}}}\right)^{2}\right], \]
where the contribution $\epsilon_{\rm t}\,\xi_{\rm e}^{2}$ comes from the definition of the magnetic moment $\mu = {\cal E}\,(1 - \epsilon_{\rm t}\,
\xi_{\rm e}^{2})/B_{\rm e}$, while the last term [with $13/12 = (39/8)\,(2/9)$] represents the anharmonic correction in the magnetic field
\eqref{eq:muBs}. We therefore have the following expression for the first-order perturbed Hamiltonian
\begin{equation}
H_{1} \;=\; \frac{m}{2}\Omega_{\mathrm{b}}^{2}s^{2}\left[ -\;\xi_{\mathrm{e}}^{2} \;+\; \frac{13}{12}\left(\frac{s}{r_{\mathrm{e}}}\right)^{2}\right]
\;+\; K_{1},
\end{equation}
where $K_{1}\equiv 9\,J_{0}^{2}/(4m\, r_{\mathrm{e}}^{2})$ is defined to satisfy 
\[ \left.\pd{K_{1}}{J_{0}}\right|_{\mu,\psi_{\mathrm{e}}} \;=\; \Omega_{\mathrm{b}}\;\frac{\xi_{\mathrm{e}}^{2}}{2} \;=\; \frac{9}{2}\;
\frac{J_{0}}{m\, r_{\mathrm{e}}^{2}}, \]
which arises from the first-order correction to the simple-harmonic bounce frequency \eqref{eq:omegab_deep}. Inserting Eqs.~\eqref{eq:J_deep} and 
\eqref{eq:s_deep} respectively for $\xi_{\mathrm{e}}\left(J_{0}\right)$ and $s\left(J_{0},\zeta_{0}\right)$ yields the first-order Hamiltonian
\begin{equation}
H_{1} \;=\; \frac{J_{0}^{2}}{mr_{\mathrm{e}}^{2}}\left(\frac{9}{4}\;-\;9\,\cos^{2}\zeta_{0}\;+\;\frac{13}{6}\;\cos^{4}\zeta_{0}\right) \;=\; 
-\;\frac{J_{0}^{2}}{mr_{\mathrm{e}}^{2}}\left(\frac{23}{16}\;+\;\frac{41}{12}\cos2\zeta_{0}\;-\;\frac{13}{48}\cos4\zeta_{0}\right),
\label{eq:deltaH}
\end{equation}
which has bounce-angle-independent and dependent parts.

\subsubsection{Anharmonic bounce action-angle transformation}

We extend our description of trapped-particle dynamics in an axisymmetric dipole magnetic field beyond the harmonic-oscillator approximation by considering the coordinate transformation $Z_{0}^{\alpha}=\left(J_{0},\zeta_{0}\right)\rightarrow Z^{\alpha}=\left(J,\zeta\right)$:
\begin{equation}
Z^{\alpha} \;=\; Z_{0}^{\alpha} \;+\; \epsilon_{\mathrm{t}}\left\{ \wh S_{1},\; Z_{0}^{\alpha}\right\}_{\|0} \;+\; \cdots,
\label{eq:ZZ_0}
\end{equation}
where the gauge function $\wh S_{1}\left(J_{0},\zeta_{0}\right)$ is designed to restore the bounce-angle independence to the bounce-center dynamics in the anharmonic approximation, and $\{,\}_{\|0}$ is defined in Eq.~\eqref{eq:PB_0}. Note that, unlike the bounce-center transformation \eqref{eq:zbar_a}, the magnetic-label flux coordinates $\left(\psi,\phi\right)$ are frozen, i.e., only the parallel motion in phase space is transformed by 
Eq.~\eqref{eq:ZZ_0}.

Using the bounce action-angle transformation \eqref{eq:ZZ_0}, we build a new bounce-angle independent Hamiltonian 
\begin{equation}
\wh H(J)  \;=\; \wh H_{0}(J) \;+\; \epsilon_{\mathrm{t}}\wh H_{1}(J) \;+\; \ldots \;=\; H_{0}(J_{0}) \;+\; \epsilon_{\mathrm{t}}\left(H_{1}(J_{0},
\zeta_{0}) \;-\frac{}{} \left\{ \wh S_{1},\; H_{0}\right\}_{\|0}(J_{0},\zeta_{0}) \right) \;+\; \cdots.
\label{eq:whH_def}
\end{equation}
From Eq.~\eqref{eq:whH_def}, we obtain the following relations at zeroth order
\begin{equation}
\wh H_{0} \;=\; \mu B_{\mathrm{e}} \;+\; J\Omega_{\mathrm{b}},
\end{equation}
and first order
\begin{equation}
\wh H_{1} \;=\; \left( \langle H_{1}\rangle_{\rm b} \;+\; \wt{H}_{1}\right) \;-\; \Omega_{\mathrm{b}}\;\pd{\wh S_{1}}{\zeta},
\end{equation}
where we have divided $H_{1}$ into a bounce-angle-independent part $\langle H_{1}\rangle_{\rm b}$ and a bounce-angle-dependent part 
$\wt{H}_{1} = H_{1} - \langle H_{1}\rangle_{\rm b}$. Since $\wh H_{1}$ must be bounce-angle-independent, we have 
\begin{equation}
\wh H_{1} \;=\; \left\langle H_{1}\right\rangle_{\mathrm{b}} \;=\; -\;\frac{23}{16}\;\frac{J^{2}}{mr_{\mathrm{e}}^{2}},
\end{equation}
while the first-order generating function $\wh{S}_{1}(J_{0},\zeta_{0})$ is defined as
\begin{equation}
\wh S_{1}(J_{0},\zeta_{0}) \;=\; \frac{1}{\Omega_{\mathrm{b}}}\int\wt H_{1}(J_{0},\zeta_{0})\;d\zeta_{0} \;=\; 
\frac{J_{0}^{2}}{24m\Omega_{\mathrm{b}}r_{\mathrm{e}}^{2}}
\left( -\;41\sin2\zeta_{0} \;+\; \frac{13}{8}\sin4\zeta_{0}\right).
\label{eq:S1hat}
\end{equation}
From Eq.~\eqref{eq:ZZ_0}, the new expressions for the bounce action-angle coordinates $\left(J,\zeta\right)$, which include anharmonic corrections, are therefore
\begin{equation}
J \;=\; J_{0} \;+\; \epsilon_{\mathrm{t}}\left\{ \wh S_{1},\;J_{0}\right\}_{\|0} \;=\; J_{0} \;+\; \epsilon_{\mathrm{t}}\,\pd{\wh S_{1}}{\zeta_{0}} \;=\; 
J_{0} \;+\; \frac{\epsilon_{\mathrm{t}}\,J_{0}^{2}}{12m\Omega_{\mathrm{b}}r_{\mathrm{e}}^{2}}\left( -\;41\cos2\zeta_{0} \;+\; \frac{13}{4}
\cos4\zeta_{0}\right),
\label{eq:Jhat}
\end{equation}
and 
\begin{equation}
\zeta \;=\; \zeta_{0} \;+\; \epsilon_{\mathrm{t}}\left\{ \wh S_{1},\;\zeta_{0}\right\}_{\|0} \;=\; \zeta_{0} \;-\; \epsilon_{\mathrm{t}}\;
\pd{\wh S_{1}}{J_{0}} \;=\; \zeta_{0} \;+\; \frac{\epsilon_{\mathrm{t}}\,J_{0}}{12m\Omega_{\mathrm{b}}r_{\mathrm{e}}^{2}}\left(41\sin2\zeta_{0} \;-\;
\frac{13}{8}\sin4\zeta_{0}\right),
\label{eq:zetahat}
\end{equation}
Equations \eqref{eq:Jhat}-\eqref{eq:zetahat} can be easily used to show that the bounce action-angle transformation \eqref{eq:ZZ_0} is canonical,
i.e., $\{\zeta, J\}_{\|0} = 1 + {\cal O}(\epsilon_{\rm t}^{2})$.

\subsubsection{Anharmonic bounce action-angle dynamics}

The new (bounce-angle-independent) bounce-center Hamiltonian now reads 
\begin{equation}
\wh H \;=\; \mu B_{\mathrm{e}} \;+\; J\,\Omega_{\mathrm{b}} \;-\; \epsilon_{\mathrm{t}}\;\frac{23}{16}\frac{J^{2}}{mr_{\mathrm{e}}^{2}},
\label{eq:Hhat}
\end{equation}
where the last term denotes the anharmonic correction. The new bounce frequency (with anharmonic correction) is therefore evaluated as
\begin{equation}
\omega_{\mathrm{b}} \;=\; \pd{\wh H}J \;=\; \Omega_{\mathrm{b}} \;-\; \epsilon_{\mathrm{t}}\frac{23}{8}\frac{J}{mr_{\mathrm{e}}^{2}} \;+\; \cdots
\;=\; \Omega_{\mathrm{b}}\left( 1 \;-\; \epsilon_{{\rm t}}\;\frac{23}{72}\;\xi_{\mathrm{e}}^{2} \;+\; \cdots\right),
\label{eq:omegab_anh}
\end{equation}
which is exactly equivalent to the direct calculation leading to
Eq.~\eqref{eq:tau_b_xi}. Figure \ref{fig:f_xi} shows the normalized bounce period $f(\xi_{\rm e}) \equiv \Omega_{\rm b}/\omega_{\rm b} \simeq [1-(23/72)\,\xi_{\mathrm{e}}^{2}]^{-1}$ (dash-dotted line) with the first-order anharmonic correction, which yields excellent agreement with the numerical result (solid line) up to about 
$\xi_{\mathrm{e}}\simeq 0.6$.

In order to prove that the new bounce action \eqref{eq:Jhat} (with anharmonic correction) is an adiabatic invariant at first order in 
$\epsilon_{\rm t}$, we calculate $\dot{J}=\left\{ J,H\right\} _{0}$ under the perturbed Hamiltonian $H=H_{0}+\epsilon_{\mathrm{t}}H_{1}$: 
\begin{equation}
\dot{J} \;=\; \dot{J_{0}} \;+\; \dot{\zeta_{0}}\left[\frac{\epsilon_{\mathrm{t}}\,J_{0}^{2}}{6m\Omega_{\mathrm{b}}r_{\mathrm{e}}^{2}}
\left(41\sin2\zeta_{0} \;-\; \frac{13}{2}\sin4\zeta_{0}\right)\right] \;+\; \cdots \;=\; {\cal O}\left(\epsilon_{\mathrm{t}}^{2}\right),
\label{eq:dotJhat}
\end{equation}
where we used the lowest-order Hamilton's equations $\dot{\zeta_{0}}=\Omega_{\mathrm{b}}$ and
\[ \dot{J_{0}} \;=\; \frac{\epsilon_{\mathrm{t}}\,J_{0}^{2}}{6mr_{\mathrm{e}}^{2}}\left( -\;41\sin2\zeta_{0} \;+\; \frac{13}{2}\sin4\zeta_{0}\right)
\;=\; -\;\epsilon_{\mathrm{t}}\,\Omega_{\rm b}\;\pd{J_{1}}{\zeta_{0}}. \]

\subsubsection{Anharmonic canonical condition}

Equations \eqref{eq:ppar_deep}-\eqref{eq:s_deep} now give the parallel canonical coordinates $(s, p_{\|})$ (with first-order corrections) as 
\begin{eqnarray}
p_{\|} & = & p_{\|0} \;+\; \epsilon_{\mathrm{t}}\left\{ \wh S_{1},\; p_{\|0}\right\}_{\|0} \;+\; \cdots \;=\; 
p_{\|0} \;+\; \pd{p_{\|0}}{J_{0}}(J-J_{0}) \;+\; \pd{p_{\|0}}{\zeta_{0}}(\zeta-\zeta_{0}) \;+\; \cdots \;=\; p_{\|0}(J,\zeta) \nonumber \\
 & \equiv & -\;\left(2m\frac{}{}J\Omega_{\mathrm{b}}\right)^{1/2}\;\sin\zeta, \label{eq:ppar_deephat} \\
s & = & s_{0} \;+\; \epsilon_{\mathrm{t}}\left\{ \wh S_{1},\; s_{0}\right\}_{\|0} \;+\; \cdots \;=\;  s_{0} \;+\; 
\pd{s_{0}}{J_{0}}(J-J_{0}) \;+\; \pd{s_{0}}{\zeta_{0}}(\zeta-\zeta_{0}) \;+\; \cdots \;=\; s_{0}(J,\zeta) \nonumber \\
 & \equiv &  \left(\frac{2J}{m\Omega_{\mathrm{b}}}\right)^{1/2}\;\cos\zeta .
\label{eq:s_deephat}
\end{eqnarray}
We therefore retain the harmonic-oscillator expressions for the parallel
coordinates, but with an anharmonic Hamiltonian. Furthermore, one
can immediately verify that 
\begin{equation}
\left\{ s,\; p_{\|}\right\}_{\|} \;=\; \pd s{\zeta}\,\pd{p_{\|}}J \;-\; \pd sJ\,\pd{p_{\|}}{\zeta} \;=\; \sin^{2}\zeta \;+\; \cos^{2}\zeta \;+\; \cdots \;=\; 1 \;+\; \mathcal{O}\left(\epsilon_{\mathrm{t}}^{2}\right),
\label{eq:shatpparhat_canonicalhat}
\end{equation}
i.e., Eqs.~\eqref{eq:ppar_deephat}-\eqref{eq:s_deephat} are canonical up to $\mathcal{O}\left(\epsilon_{\mathrm{t}}^{2}\right)$.

\subsubsection{Anharmonic-corrected bounce-averaged drift frequency}

Lastly, we can also obtain anharmonic corrections to the bounce-average drift frequency \eqref{eq:omega_d_xi} as follows. First, we express the magnetic-field strength \eqref{eq:B_sigma} at constant ${\cal M} = \psi\,r_{\rm e} = B_{\rm e}\,r_{\rm e}^{3}$ and $s$ as
\[ B(\psi, s) \;=\;  \frac{\psi^{3}}{{\cal M}^{2}} \;+\; \epsilon_{\rm t}\;\frac{9}{2}\;s^{2}\; \frac{\psi^{5}}{{\cal M}^{4}} \;+\; \cdots. \] 
Next, the flux derivative $\partial B/\partial\psi$ (at constant $s$) is evaluated as
\begin{equation}
\pd B{\psi} \;=\; \frac{3}{r_{\mathrm{e}}^{2}}\left[\; 1 \;+\; \epsilon_{\rm t}\;\frac{15}{2}\left(\frac{s}{r_{\mathrm{e}}}\right)^{2} \;+\; 
\cdots \;\right],
\label{eq:dB_dpsi_xi}
\end{equation}
where we have retained the anharmonic first-order correction. Next, by using the expression \eqref{eq:s_deephat} for the parallel spatial coordinate $s(\zeta)$, the bounce-average drift frequency \eqref{eq:omega_d_xi} becomes
\begin{eqnarray}
\omega_{\rm d} & = & \frac{3c\,{\cal E}}{q\,B_{\rm e}\,r_{\rm e}^{2}} \left(1 - \epsilon_{\rm t}\;\xi_{\rm e}^{2}\right) \left( 1 \;+\; \epsilon_{\rm t}\;\frac{15}{18}\;\xi_{e}^{2} \;+\; \cdots \right) \;=\; \Omega_{\rm d} \left[\; 1 \;-\; \epsilon_{\rm t} \left(1 - \frac{15}{18}\right)\;\xi_{\rm e}^{2} \;+\; \cdots \;\right] \nonumber \\
 & \simeq & \Omega_{\rm d} \left( 1 \;-\; \epsilon_{\rm t}\;\frac{\xi_{\rm e}^{2}}{6} \right),
\label{eq:omega_d_anh}
\end{eqnarray}
where we have combined the definition $\mu\,B_{\rm e} = {\cal E}\,(1 - \epsilon_{\rm t}\,\xi_{\rm e}^{2})$ with the bounce-averaged expression for Eq.~\eqref{eq:dB_dpsi_xi}. Figure \ref{fig:g_xi} shows the normalized bounce-average drift period $g(\xi_{e}) \equiv \Omega_{\rm d}/\omega_{\rm d} 
\simeq [1 - (1/6)\,\xi_{\rm e}^{2}]^{-1}$ (dash-dotted line) with the first-order anharmonic correction, which yields excellent agreement with the numerical result (solid line) up to about $\xi_{\rm e} \simeq 0.6$.

\subsection{Bounce-center coordinates}

The first-order generating vector field for the bounce-center transformation \eqref{eq:bct_def} is calculated using the expressions for the deeply-trapped action-angle coordinates. Equations \eqref{eq:G1_a_def} and \eqref{eq:phidot_dip} are used, with the lowest-order expressions 
$\omega_{\rm b}^{-1}\,d\zeta = m\,ds/p_{\|}$, to obtain the expression for $G_{1}^{\phi}$:
\begin{equation}
G_{1}^{\phi} \;=\; -\;\frac{c}{q}\;a\,p_{\|} \;+\; \frac{c\mu}{q\omega_{\rm b}}\int\left(\left\langle \pd B{\psi}\right\rangle_{\mathrm{b}} \;-\; 
\pd B{\psi}\right)d\zeta,
\label{eq:G1phi_deep}
\end{equation}
where, using Eq.~\eqref{eq:dB_dpsi_xi}, we find
\[ \left\langle \pd B{\psi}\right\rangle_{\mathrm{b}} \;-\; \pd B{\psi} \;=\; -\;\epsilon_{\rm t} \frac{45\,J\,\cos 2\zeta}{2\,m\,\Omega_{\rm b}
r_{\rm e}^{2}} \;+\; \cdots. \]
In the lowest-order deeply-trapped approximation, where only the first term in Eq.~\eqref{eq:G1phi_deep} remains, we find
\begin{equation}
G_{1}^{\phi} \;=\; 3\;\frac{cJ}{q\psi_{\rm e}}\;\sin 2\zeta \;+\; \cdots \;=\; \frac{J\,\Omega_{\rm d}}{{\cal E}}\;\sin 2\zeta \;+\; \cdots,
\label{eq:G1phiJhatzetahat}
\end{equation}
where we used Eq.~\eqref{eq:a_deep} for $a$ and Eq.~\eqref{eq:Omega_d_def} for $\Omega_{\rm d}$. Here, we easily verify that $\langle G_{1}^{\phi}
\rangle_{\mathrm{b}}=0$.

It was shown in Section~\ref{sec:bc_general} that the flux coordinate $\psi$ is a guiding-center invariant in axisymmetric magnetic-dipole geometry, which ensures that $G_{1}^{\psi}=0$. Hence, the bounce-center position $\ov{{\bf X}}$ is a purely toroidal shift of the bounce-averaged particle position 
$\left\langle \mathbf{X}\right\rangle _{\mathrm{b}}$, expressed as
\begin{equation}
\ov{{\bf X}} \;=\; \left\langle \mathbf{X}\right\rangle_{\mathrm{b}} \;+\; \epsilon\; G_{1}^{\phi}\left(\psi,J,\zeta\right)\;\partial
{\bf X}/\partial\phi.
\label{eq:Xbc}
\end{equation}
We see that the general bounce-center transformation is greatly simplified
in the axisymmetric dipole geometry, and is consistent with the purely
toroidal nature of the magnetic gradient and curvature drifts in this
geometry.

\section{\label{sec:summary}Summary}

We now summarize the main results of the present work. The bounce-center transformation that transforms the guiding-center phase-space Lagrangian 
\eqref{eq:psl_alphabeta} into the bounce-center phase-space Lagrangian \eqref{eq:psl_bc} relies on the existence of expressions for the parallel canonical coordinates $s(J,\zeta)$ and $p_{\|}(J,\zeta)$ that are required to satisfy the canonical identity \eqref{eq:canonical_id}. In general magnetic geometry, this canonical identity is taken as an axiom for the bounce-center transformation \cite{RGL_82a}, since the identity is difficult to prove directly beyond the harmonic-oscillator approximation. For practical applications in axisymmetric tokamak geometry \cite{TSH_2009}, for example, it is a common practice to use expressions for $s(J,\zeta)$ and $p_{\|}(J,\zeta)$ in the deeply-trapped approximation. The primary purpose of the present work was, therefore, to investigate the canonical identity \eqref{eq:canonical_id} beyond the harmonic-oscillator approximation in an axisymmetric magnetic geometry simple enough to allow explicit analytical results. The axisymmetric dipole geometry was chosen for its great simplicity (e.g., all confined particles are trapped) as well as its universality in space plasmas \cite{Parks} and its applications in some recent laboratory experiments \cite{LDX}.

By using elegant Lie-transform Hamiltonian perturbation methods, we have been able to obtain explicit expressions (with harmonic and anharmonic contributions) for the parallel canonical coordinates $s(J,\zeta)$ and $p_{\|}(J,\zeta)$ in axisymmetric dipole geometry, which satisfy the canonical identity \eqref{eq:canonical_id}. By including anharmonic corrections to the deeply-trapped particle dynamics, we were also able to obtain explicit expressions for the bounce frequency \eqref{eq:omegab_anh} and the bounce-averaged drift frequency \eqref{eq:omega_d_anh} that yielded excellent agreement (up to $\xi_{\rm e} \simeq 0.6$) with the exact numerical results shown in Figs.~\ref{fig:f_xi} and \ref{fig:g_xi}, respectively, as well as standard numerical fits \cite{Hamlin}.

Future work will focus on the bounce-center transformation in axisymmetric tokamak geometry \cite{Brizard_oagcfp} beyond the deeply-trapped approximation, for which the poloidal flux $\psi$ is no longer an invariant, but is replaced with the drift invariant $\psi^{*} \equiv \psi - \rho_{\|}\,B_{\phi}$, where $\rho_{\|} = p_{\|}/(m\omega_{\rm g})$ and $B_{\phi}$ denotes the toroidal covariant component of the tokamak magnetic field.

\begin{acknowledgments}
One of us (AJB) would like to acknowledge the kind hospitality of
the Institut de la Recherche sur la Fusion par confinement Magn\'{e}tique
at CEA Cadarache. This work, supported by the European Communities
under the contract of Association between EURATOM and CEA, was carried
out within the framework of the European Fusion Development Agreement.
The views and opinions expressed herein do not necessarily reflect
those of the European Commission.
\end{acknowledgments}

\end{document}